
\input jnl.tex

\vglue 1. truein
\title
{
Peculiar Velocities and Microwave Background Anisotropies
from Cosmic Strings
}
\author
{Leandros Perivolaropoulos}
\affil
{
Harvard-Smithsonian Center for Astrophysics,
60 Garden Street,
Cambridge, MA 02138.
}
\author
{Tanmay Vachaspati}
\affil
{
Tufts Institute of Cosmology,
Department of Physics and Astronomy,
Tufts University, Medford, MA 02155
}
\vskip .5cm
{\rm Subject headings}: {\rm cosmology-galaxies:clustering}
\abstract
\singlespace
Using an analytic model we show that the predictions of the cosmic string model
for the peculiar velocities and the microwave background (MBR) anisotropy
depend on similar combinations of string evolution parameters.
Normalizing from the COBE detection of MBR anisotropy, and for certain
reasonable values of string network evolution parameters,
we find that the magnitude of predicted velocity flows is in good agreement
with observations on small scales but is inconsistent with
observations on large scales ($ > 50 h^{-1}$Mpc).
The Cosmic Mach Number obtained from the cosmic string scenario is found to
depend on a single network evolution parameter and is
consistent with observations on scales  $5h^{-1}$Mpc to $20h^{-1}$Mpc.

\endtopmatter

The cosmic string model for large-scale structure formation
with hot dark matter promises to be
successful in many ways
(Vachaspati T. \& Vilenkin A. 1991; Vollick D. N. 1992;
Albrecht A. \& Stebbins A. 1993).
The scenario might automatically account for large-scale filaments
and sheets (Vachaspati T. 1986; Stebbins A. et al. 1987;
Vachaspati T. \& Vilenkin A. 1991;
Perivolaropoulos L., Brandenberger R. \& Stebbins A. 1990;
Hara T. \& Miyoshi S. 1990),
galaxy formation at epochs of $z \sim 2-3$ and
galactic magnetic fields (Vachaspati T. \& Vilenkin A. 1991;
Vachaspati T. 1992b).
It can also provide (Vachaspati T. 1992a; Brandenberger R. et al. 1987)
large-scale peculiar velocities (Dressler A. et. al. 1987; Burstein D. et. al.
1983; Peebles P. J. E. \& Silk J. 1990)
and is consistent (Bennett D. P., Stebbins A. \& Bouchet F. R. 1992;
Perivolaropoulos L. 1993a,b) with the measured anisotropy
of the microwave background radiation (Smoot G. et. al. 1992).

Even though the cosmic string model involves a single free parameter (the
mass per unit length of the string), its predictions
depend also on our understanding of the evolution of the string network.
The main features of this evolution can be encoded into a set of parameters
which have been constrained by recent detailed numerical
simulations (Bennett D. P. \& Bouchet F. R. 1988;
Allen B. \& Shellard E. P. S. 1990) (see also papers by Albrecht \& Turok in
(Gibbons G. W., Hawking S. W. \& Vachaspati T. 1990)).
Therefore, in order to test the cosmic string model, it is
important to obtain expressions of the  model's observational predictions in
terms of these parameters and then use observations to fix their values. The
consistency of the model can then be tested by comparing the fixed values with
numerical simulations of string evolution and with constraints from different
observations.

In this letter we show that the predictions of the peculiar velocities
and MBR anisotropies resulting from cosmic strings depend on roughly the same
combination of undetermined parameters  which we call $\alpha$. The prediction
of peculiar velocities also depends on parameters denoted
$\xi$ and $f_{5/4}$, denoting the curvature radii of strings and the
Newtonian interaction of wiggly strings with matter. On the other hand,
the microwave background predictions are independent of $f_{5/4}$ (since there
is no Newtonian interaction of strings with photons). By combining the two
predictions and comparing with observations, we can impose constraints on
these parameters, thus providing an interesting test for the cosmic
string model. One outcome is that the string scenario cannot account
for the observed peculiar velocities on all the different scales. With
the most suitable choice of parameters, there is good agreement with
observations on small scales but inconsistency with the observations
on large scales. This potential problem of cosmic strings however, is also
encountered in several other models (Brandenberger R. et. al. 1987) including
the standard CDM model.

It is not difficult to guess that the peculiar velocities and the
microwave background anisotropy should depend on roughly the same combination
of network parameters. The peculiar velocities arise because
with every string sweeping across the horizon, any fixed volume of
matter experiences an impulse towards the wake of the string. At the
same time, the anisotropy arises because the
string gives an impulse to the photons coming
towards us from the surface of last scattering. The peculiar velocity
that we observe today, would be the sum of all string impulses with
some suitable growth factors while the anisotropy would be due to
the sum of the string impulses on the photons from the surface of
last scattering. Therefore both processes depend on evaluating
the magnitude of each impulse, the number of such impulses and, in
the case of peculiar velocities, the growth of the velocity following
the impulse. Since the calculations are so similar, the dependences
on the string network parameters are also very similar.

Let us first apply the multiple impulse approximation (Vachaspati T. 1992a;
Perivolaropoulos L. 1993a,b)
to obtain the mean and the standard deviation of
cosmic string induced peculiar velocities.
According to this approximation, the time between $t_{eq}$ (the time of equal
matter and radiation) and the present
($t_p$) is divided into a set of Hubble time-steps $t_i$ such that
$t_{i+1}=2 t_i$. At every Hubble-step
$t_i$, each long string gives a velocity impulse to matter within a
distance $\xi t_i$ from it, where $\xi t_i$ is the radius of curvature of
the string, and, by causality $\xi \leq 1$.
It is straightforward to count the total number of string induced impulses
for each comoving scale $Lh^{-1}Mpc$. By including the growth factor
$({t_p\over t_n})^{1/3}$ for the n$^{th}$ impulse, the velocity ${\vec
v}(t_p,L)$ of the scale $L$ at $t_p$ may be written as:
$$
{\vec v}(t_p,L)={2\over 5}\Delta v
\sum_{n=0}^{N_L} \sum_{m=1}^{n_s (L)} {\hat k}_{n m} ({t_p \over t_n})^{1/3}
\eqno (1)
$$
where
$$
\Delta v=5 \pi G \mu v_s \gamma_s f_{5/4}
\eqno (2)
$$
and the indices $n$, $m$ count Hubble steps and long strings within a Hubble
volume respectively.
In (1), (2) and in what follows we use the following notation:
$N_L$ is the number of Hubble time steps during which a volume of
comoving size $L$ experiences coherent string impulses,
$n_s (L)$ is the number of string impulses that the volume
experiences per Hubble time,
$v_s$ is the string velocity and $\gamma_s$ the corresponding Lorentz
factor, $(G\mu_0)_6$ is the {\it bare} string tension in units of $10^{-6}$,
$$
f_{5/4}= {4\over 5} \left [ 1+{1\over 2}
        {1\over {(v_s \gamma_s)^2}} (1- {T\over \mu}) \right ]
\eqno (3)
$$
and $T$, $\mu$ are the
renormalized string tension and string mass density.
Simulations (Allen B. \& Shellard E. P. S. 1990;
Bennett D. P. \& Bouchet F. R. 1988)
indicate $T \sim 0.7 \mu_0$ and $\mu \sim 1.4 \mu_0$ where
$\mu_0$ is the bare string tension. The expression for $f_{5/4}$ in
eq. (3) can be thought of in the following way: if we multiply $f_{5/4}$
by $\Delta v$, the first term is the induced velocity due to the
conical deficit of the string metric while the second term (proportional
to $1/v$) is due to the Newtonian potential of the wiggly string.
In what follows, we shall not evaluate $f_{5/4}$ using the simulation
results but treat it like a free parameter.

If we imagine a volume of comoving size $L$, the whole volume will get
a coherent impulse due to strings moving outside the volume but within a
distance equal to the curvature radius of the string. If a string
passes through the volume, it will not cause a coherent motion of the
volume and so we neglect the effect of such strings. On the other hand,
a wiggly string which is distant from the volume would still
have some Newtonian gravitational effect for $\xi<1$ (expressed by
second term in $f_{5/4}$)
on the volume and, strictly speaking, we should take the distant
strings into account with an effectively smaller value of $f_{5/4}$.
(We thank Alex Vilenkin for emphasizing this point.)
However, since we are keeping the number of strings ($n_s$) and $f_{5/4}$
as free parameters, the exact values do not concern us and we should
view $f_{5/4}$ as an average value that accounts for all the strings.
In addition, for $\xi \sim 1$, the curvature radius is comparable to the
horizon and our method automatically accounts for all strings.
The number of strings that give coherent impulses to the volume can now
be found: it is simply the number of strings within the horizon multiplied
by the probability that the string passes within a distance $\xi t_n$
from the volume of size $L$. Therefore,
$$
n_s (L) = n_s \biggl ( {\xi t_n  - {{L(t_n )}} \over {t_n}} \biggr )
\eqno (4)
$$
provided $L(t_n ) < \xi t_n$ and $n_s (L) = 0$ otherwise. Here
$L(t_n ) =  L ( t_n / t_p )^{2/3}$ is the physical
size of the volume at the $n^{th}$ Hubble step and $n_s$ is the number
of strings within the horizon.

Now we square (1) and take averages with the assumption that the $\hat k$
vectors are random. That is,
$$
< {\hat k}_{nm} \cdot {\hat k}_{n'm'} > = \delta_{nn'} \delta_{mm'}
\eqno (5)
$$
This gives (Vachaspati T. 1992a) the present
average (rms) peculiar velocity magnitude on a comoving scale $Lh^{-1}Mpc$:
$$
{\bar v}(t_p , L \ge \xi t_{eq} ) =
                         44 \alpha \xi f_{5/4} L_{100}^{-1} \  km/s
\eqno (6)
$$
where
$$
\alpha=\sqrt{n_s \xi} (G\mu_0)_6 (v_s \gamma_s) \  ,
\eqno (7)
$$
$L_{100}\equiv L/ (100 h^{-1} Mpc )$ and we have taken $N_L >> 1$.
The result (6) is valid on scales $Lh^{-1}$ larger than the curvature
radius of strings at $t_{eq} = 10h^{-2}Mpc$ because in deriving it
we assumed $L(t_0 ) > \xi t_{eq}$.
For scales $L < \xi t_{eq}$ we get
$$
{\bar v}(t_p , L \le \xi t_{eq} ) =
          437 \left [ 1 + 2.9 \left ( 1 - {L \over {\xi t_{eq}}}
                   \right ) \right ]^{1/2} \alpha f_{5/4} h \ km/s \ .
\eqno (8)
$$

The dispersion $\sigma (L)$ of ${\bar v}(L)$ may also
be obtained (Vachaspati T. 1992a) as:
$$
\sigma(t_p , L)=0.36\hskip 0.1cm {\bar v}(t_p , L) \ .
\eqno (9)
$$

As discussed above, the free parameter $\alpha$ involved in the peculiar
velocity calculation can be pinned down by demanding that the cosmic
string model be consistent with the recent detection of MBR anisotropy by
COBE. The multiple impulse approximation can be used (Perivolaropoulos L.
1993a)
to express the string induced MBR anisotropies in terms of $\alpha$. In this
case the approximation is realized as follows:  A given photon
experiences the impulses due to the various strings along its way from the
last scattering surface to us.
Each impulse can either red shift or blue shift the photon,
depending on the various possible directions of the string velocity, photon
momentum and string orientation. The temperature variations at some fixed
angular separation will be correlated only when the photon beams at
that angular separation will experience the same impulse.
The correlation function is thus obtained by counting the number
of common impulses on each angular scale $\theta$. In this case, due to the
lack of a growth factor, the dependence on $\xi$ (the string curvature
radius) comes only as a square root dependence in counting the number
of strings that come within a radius of curvature distance from the
photon.
{}From the
correlation function it is straightforward to obtain the rms temperature
fluctuation for angular beam separation equal to $60^\circ$ and smoothing on
a scale of $10^\circ$ as in the COBE experiment. The result is
(Perivolaropoulos L. 1993a):
$$
\biggl (
{{\Delta T} \over T} \biggr )_{rms} = 1.6 \hskip 0.1cm \alpha \times 10^{-5}
\eqno (10)
$$
where $\alpha$ is the same product of parameters as in (7).

Now the COBE observations fix the rms temperature fluctuations to be
$({{\Delta T}\over T})_{rms}=(1.1\pm 0.2)\times 10^{-5}$ which when
inserted in (6) gives
$$
\alpha \simeq 0.7 \pm 0.2
\eqno (11)
$$

It is of interest to use the values of evolution parameters
$n_s$, $v_s \gamma_s$ and $\xi$ obtained in numerical simulations to
extract from the value of $\alpha$ in (11), a rough value of the single
free parameter of the string model $G\mu_0$. From
(Allen B. \& Shellard E. P. S. 1990)
we have $n_s \simeq 10$, $v_s \simeq 0.15$
and $\xi \simeq 0.7$. Using these values in (7) we obtain (Perivolaropoulos L.
1993a): $$
G\mu_0 =(1.8\pm 0.5) \times 10^{-6}
\eqno (12)
$$
which is the value required from galaxy and large scale structure
calculations (Perivolaropoulos L., Brandenberger R. \& Stebbins A. 1990;
Turok N. \& Brandenberger R. 1986; Sato H. 1986; Stebbins A. 1986). (The
value of $G\mu$ given in (Perivolaropoulos L. 1993a) refers to the {\it
effective} $G\mu$ which includes the string small scale structure and is
therefore slightly different from (12)). The
same value for $G\mu$ has been obtained in
(Bennett D. P., Stebbins A. \& Bouchet F. R. 1992) by using
the numerical simulations of (Bennett D. P. \& Bouchet F. R.1988;
Bouchet F. R., Bennett D. P. \& Stebbins A. 1988) to simulate
the MBR sky and compare with the COBE results. This is an additional test of
consistency for our results.

Now, using (11) in (6) and taking into account a $1\sigma$ spread
(eq. (9)) we get
$$
{\bar v}(t_p, L \ge \xi t_{eq}) =
      (31 \pm 11) \hskip 0.1cm \xi f_{5/4} L_{100}^{-1}\hskip 0.2cm km/s
\eqno (13)
$$
$$
{\bar v}(t_p , L \le \xi t_{eq} ) =
               (306 \pm 110) \left [ 1 +
                  2.9 \left ( 1 - {L \over {\xi t_{eq}}}
                           \right ) \right ]^{1/2} f_{5/4} h \ km/s \ .
\eqno (14)
$$
The dependence of ${\bar v}$ on the length scale is plotted in Fig. 1
for $f_{5/4}=1.8$, $\xi = 1$ and $h=0.5$.
The observations (Burstein D. et. al. 1983;
Collins C., Joseph R. \&  Robertson N. 1986; Dressler A. et. al. 1987;
Groth J.,Juszkiewicz R. \& Ostriker J. 1989)
are also indicated with their error bars.

The figure shows that the peculiar velocities from cosmic strings are in
good agreement with observations on small scales but not on large scales.
The inconsistency is not an artifact of our choice of parameters.
For example,
we could take $\xi = 1$ (the largest permissible value), $h=0.5$ (the
smallest value of $h$) and
$f_{5/4} = 7$ so that the observation on $60 h^{-1}$Mpc is barely
accommodated within the $1\sigma$ prediction of the string model.
But then, the predicted $1\sigma$ velocity on scales of $8 h^{-1}$Mpc
range from $1135$km/s to $2409$km/s. These velocities are clearly
inconsistent with observations. In other words, no choice of parameters
will allow us to explain the velocity observations on all scales.

We now consider the Cosmic Mach Number (CMN) induced due to strings.
The CMN has been shown to be independent of various uncertainties such
as biasing and hence may be a valuable statistic in comparing theory with
observation (Ostriker J. \& Suto Y. 1990). As we shall see, the string induced
CMN only depends on the parameter $\xi$ and for reasonable values of $\xi$ the
agreement between the string predictions and observations on small scales is
quite good. Hence we believe that cosmic strings can successfully explain the
observations on small scales but fail in explaining the large scale data.

The Cosmic Mach Number $M(L,a)$ is defined (Ostriker J. \& Suto Y. 1990) as
the ratio of the rms velocity ${\bar v}(t_p , L)$
of a volume of size
$Lh^{-1}$ over the dispersion of the velocity field smoothed on a
scale {\it a}, in the rest frame and within the same volume:
$$
M(L,a) \equiv
{{{{\bar v}(t_p,L)}}\over
{\sqrt{<({\vec v}(t_p,a) - {\vec v}(t_p,L))^2>}}}
\eqno (15)
$$
Inserting (8) for the peculiar velocities gives us (for
$a, L < \xi t_{eq}$),
$$
M(L, a) = \sqrt{{1.35 \xi t_{eq} - L} \over {L-a}} \ .
\eqno (16)
$$
If $a < \xi t_{eq} < L$, we get,
$$
M(L, a) = \sqrt{ {0.7 \xi ^2 t_{eq} ^2 } \over
                {2.7 L^2 - 2 a L^2 /\xi t_{eq} - 0.7 \xi^2 t_{eq} ^2}} \ .
\eqno (17)
$$
Observations on scales larger than $5h^{-1}Mpc$ indicate
(Ostriker J. \& Suto Y. 1990) that
$$\eqalign{
M(L=8h^{-1}Mpc,a=5h^{-1}Mpc)&=2.2\pm 0.5 \cr
M(L=18h^{-1}Mpc,a=5h^{-1}Mpc)&=1.3\pm 0.4
\quad .
\cr}
\eqno (18)
$$
while the corresponding predictions of the cosmic string scenario
obtained from (16) for $\xi = 1$, are
$$\eqalign{
M_s(L=8h^{-1}Mpc,a=5h^{-1}Mpc)&= 2.5\cr
M_s(L=18h^{-1}Mpc,a=5h^{-1}Mpc)&= 0.8
\quad .
\cr}
\eqno (19)
$$
The CMN predictions are smaller for smaller values of $\xi$. For
example, if $\xi = 0.7$, one gets 1.9 and 0.5 in eq. (19). Simulations
indicate $\xi \simeq 0.7$, and so we can say that
the cosmic string predictions are in rough agreement with the CMN
observations. If, however, the simulations can pin down $\xi$ better,
the CMN observations would be a powerful test for the cosmic string
scenario.

The success of the string scenario with the observations on small scales
suggests that we should choose our parameters so that both the peculiar
velocity magnitudes and the CMN are in good agreement on these scales.
That is, we choose $\xi = 1$, $f_{5/4} = 1.8$ and $h= 0.5$. With this
choice of parameters, the peculiar velocity on $60 h^{-1}$Mpc is
$89 \pm 30$km/s and is off from the observations of roughly
$600$km/s by a factor of about 7.
This gives us an idea of the magnitude of the problem.

We are unable to give a spread on the predicted values of the CMN because
our analytic model does not take the correlations of velocity impulses
on the large and small scales into account.
These correlations are important for the
CMN because the dispersion occurring in the denominator of eq. (15) is
the dispersion of the smaller scale volumes (size $a$) within a bigger
volume (size $L$). For example, it is quite possible that if the bigger
volume has a larger than average velocity, then the smaller volumes
will also have a larger than average velocity. Such correlations cannot be
taken into account in our model since we can only find the average
velocities of the small and big volumes. It would also be wrong to
include the spread in peculiar velocities (eq. (9)) and simply propagate
these into the CMN because the calculated spread includes
all volumes of size $a$ and not just those contained within some larger
volume of size $L$.

The prediction of the cosmic string model is in better agreement with
observations than adiabatic CDM
which predicts (Ostriker J. \& Suto Y. 1990) $M(L,a=5h^{-1}Mpc)$ to be well
below 1 on both $8h^{-1} Mpc$ and $18h^{-1} Mpc$.

In conclusion, we have used the multiple impulse approximation to identify two
new features of the cosmic string model valid with both hot and cold dark
matter:

a) The Cosmic Mach Number predicted by the model is consistent with
observations on scales $5-20h^{-1}$Mpc.

b) The magnitude of the predicted large scale velocity flows normalized
by the COBE detection of MBR anisotropy is consistent with observations on
small scales for reasonable values of the string evolution parameters
but the large scale peculiar velocities cannot be explained with these
parameters.

The potential problem pointed out in b) is also encountered in other theories
like standard CDM where a similar $L^{-1}$ scaling of the magnitude of large
scale velocity flows has been shown (Kaiser N. 1983;
Vittorio N. \& Silk J. 1985; Vittorio N. \& Turner M. S. 1987;
Kolb E. W. \& Turner M. S. 1990).

\medskip

\noindent {\it Acknowledgements}:
We thank Albert Stebbins for initiating this work by pointing out that
there may be a problem with large-scale peculiar velocity flows in the
cosmic string scenario. We are grateful to Robert Brandenberger and
Alex Vilenkin for comments.
This work was supported by a CfA Post-Doctoral Fellowship (L.P.) and  by the
National Science Foundation (T.V.).


\references

Albrecht A. \& Stebbins A. 1993. Phys. Rev. Lett. {\bf 69}, 2615.

Allen B. \& Shellard E. P. S. 1990. Phys. Rev. Lett. {\bf 64},
119-122. See also the paper by Shellard \& Allen in
(Gibbons G. W., Hawking S. W. \& Vachaspati T. 1990).

Bennett D. P. \& Bouchet F. R.1988. Phys. Rev. Lett. {\bf{60}},
257-260 (1988). See also the papers by Bennett and
Bouchet in (Gibbons G. W., Hawking S. W. \& Vachaspati T. 1990).

Bennett D. P., Stebbins A. \& Bouchet F. R. 1992.  Ap. J. Lett. {\bf 399},
L5.

Bouchet F. R., Bennett D. P. \& Stebbins A. 1988.  Nature {\bf
335}, 410.

Brandenberger R., Kaiser N., Shellard E. P. S., Turok N. 1987. Phys.Rev.
{\bf D36}, 335.

Burstein D. et. al.1983. in ``Galaxy Distances and
Deviations from the Hubble Flow'', eds. B. Madore \& R. Tully
(Dordrecht: Reidel, 1983)

Collins C., Joseph R. \&  Robertson N. 1986.
Nature {\bf 320}, 506-508.

Dressler A., Faber S. M., Burstein D., Davies R. L., Lynden-Bell D.,
Terlevich R. J. \& Wegner G. 1987.  Ap. J. Letters, {\bf 313}, L37.

Hara T. \& Miyoshi S. 1990. Prog. Theor. Phys. (Japan) {\bf{81}},
1187-1197.

Gibbons G. W., Hawking S. W. \& Vachaspati T. 1990. ``The
Formation and Evolution of Cosmic Strings'', Cambridge University Press.

Groth J., Juszkiewicz R. \& Ostriker J. 1989.  Ap.J. {\bf 346} 558.

Kaiser N. 1983. Ap. J. {\bf 273}, L17.

Kolb E. W. \& Turner M. S. 1990. ``The Early Universe'',
Addison Wesley.

Ostriker J. \& Suto Y. 1990. {\it Ap.J.} {\bf 348}, 378.

Peebles P. J. E. \& Silk J. 1990. Nature {\bf 346},
233-239.

Perivolaropoulos L. 1993a. Phys. Lett. {\bf B298 }, 305.

Perivolaropoulos L. 1993b. {\it On the Statistics of CMB Fluctuations
Induced by Topological Defects}, submitted to Phys. Rev. {\bf D}.

Perivolaropoulos L., Brandenberger R. \& Stebbins A. 1990.
Phys. Rev. {\bf D41}, 1764.

Sato H. 1986. {\it Prog. Theor. Phys.} {\bf 75}, 1342.

Smoot G. et. al. 1992. {\it Ap. J. Lett.} {\bf 396}, L1.

Stebbins A. 1986. {\it Astrophys. J. Lett.} {\bf 303}, L21.

Stebbins A. et al. 1987. Ap. J. {\bf 322}, 1.

Turok N. \& Brandenberger R. 1986.
Phys. Rev. {\bf D33}, 2175.

Vachaspati T. 1986. Phys. Rev. Lett. {\bf 57}, 1655.

Vachaspati T. 1992a. Phys. Lett. {\bf B282}, 305.

Vachaspati T. 1992b. Phys. Rev. D{\bf 45}, 3487.

Vachaspati T. \& Vilenkin A. 1991. Phys. Rev. Lett. {\bf 67},
1057-1061.

Vittorio N. \& Silk J. 1985. Ap. J. {\bf 293}, L1.

Vittorio N. \& Turner M. S. 1987. Ap. J. {\bf 316}, 475.

Vollick D. N. 1992. Phys. Rev. D{\bf 45}, 1884.

\endreferences


\head {Figure Captions}

\item {1.} The predicted peculiar velocity versus the scale
lies in the shaded area
between the two shown curves within a $1\sigma$ error including the standard
deviation of ${\bar v}$. The data points with error bars are the observations.
The parameter values $f_{5/4}=1.8$, $\xi=1$, $h=0.5$ were used.
\vfill
\eject

\endjnl

\end